\documentclass[aps,prb,preprint]{revtex4}
\usepackage{graphicx}
\usepackage{dcolumn}
\usepackage{bm}

\begin{document}


\title{Effective masses and complex dielectric function of cubic $HfO_2$}

\author{J. C. Garcia, L. M. R. Scolfaro, J. R. Leite, A. T. Lino}

\address{Instituto de F\'\i sica, Universidade de S\~ao Paulo, C. P. 66318, 05315-970
S\~ao Paulo, SP, Brazil}

\author{V. N. Freire, G. A. Farias}

\address{Departamento de F\'\i sica, Universidade Federal do
Cear\'a, C. P. 6030, 60455-900 Fortaleza, CE, Brazil}

\author{E. F. da Silva Jr.}

\address{Departamento de F\'\i sica, Universidade Federal de
Pernambuco, 50670-901 Recife, PE, Brazil}

\date{September 21, 2004)}



\begin{abstract}

The electronic band structure of cubic $HfO_2$ is calculated using
an {\it ab initio} all-electron self--consistent linear augmented
plane-wave method, within the framework of the local-density
approximation and taking into account full-relativistic
contributions. From the band structure, the carrier effective
masses and the complex dielectric function are obtained. The
$\Gamma$-isotropic heavy and light electron effective masses are
shown to be several times heavier than the electron tunneling
effective mass measured recently. The imaginary part of the
complex dielectric function $\epsilon_2(\omega)$ is in good
agreement with experimental data from ultraviolet spectroscopic
ellipsometry measurements in bulk yttria-stabilized $HfO_2$ as
well as with those performed in films deposited with the tetrakis
diethylamido hafnium precursor for
energies smaller than 9.5\thinspace eV. \\
PACS[78.20.Ci, 71.18.+y,71.20.-b]

\vspace{3cm}

\noindent Contact author: scolfaro@macbeth.if.usp.br

\end{abstract}


\maketitle


Aggressive scaling of the gate lengths and equivalent gate oxide
thickness following the {\it International Technology Roadmap for
Semiconductors} (ITRS) \cite{ITRS} is forcing the replacement of
silicon dioxide as a gate dielectric by high-dielectric constant
(high--$\kappa$) materials to reduce leakage currents
\cite{Packan1}, and meet requirements of reliability
\cite{Stathis2}. $Si$-related band offsets, permittivity,
dielectric breakdown strength, interface stability and quality
with silicon, and the carrier effective masses are key aspects of
the new (high--$\kappa$) oxides that, together with costs of the
fabrication processes \cite{Stoneham3}, have to be considered in
extending Moore's law \cite{Moore4} to an equivalent oxide
thickness below 10\thinspace nm.

Cubic hafnium oxide ($HfO_2$) is an important candidate for
$SiO_{2}$ replacement as gate dielectric material. It has a
dielectric constant $\varepsilon_{HfO_2}(0) \simeq 25$ at
$300\thinspace K$, which is about six times higher than that of
silicon dioxide, but more than an order of magnitude smaller than
that of cubic $SrTiO_3$, $\varepsilon_{SrTiO_3}(0) \simeq 300$, a
competing high--$\kappa$ oxide. However, $HfO_2$ has a conduction
band offset $\Delta E_{c,{HfO_2}}\sim 1.5-2.0 \thinspace eV$ with
respect to silicon, which is more than one order of magnitude
higher than that of cubic $SrTiO_3$, $\Delta E_{c,{SrTiO_3}}\sim
0.1 \thinspace eV$ \cite{Afanasev5}. The high dielectric constant
and tunneling barrier (with respect to silicon) of $HfO_2$,
together with a possible adaptation of the device production line
to hafnium oxide thin film growth, make its candidate stronger for
$SiO_{2}$ replacement as gate dielectric material. Recently,
improvements in the growth techniques of thin $HfO_2$ films lead
to a gate dielectric with $0.5\thinspace nm$ equivalent oxide
thickness \cite{Harris6}.

A first-principles study of the structural, vibrational, and
lattice dielectric properties of bulk hafnium oxide in the cubic,
tetragonal, and monoclinic phases has been performed
\cite{Vanderbilt7}, as well as a theoretical evaluation of some
properties of thin $HfO_2$ films on Si(001) \cite{Fiorentini8}.
However, few results concerning band-structure calculations have
been published within the non-relativistic approach, or in the
relativistic approach including spin-orbit interaction effects
\cite{Boer9,Peacock10,Demkov11}. The values of the $HfO_2$ carrier
effective masses, which are fundamental to the modeling of
tunneling currents through $HfO_2$ gate dielectrics
\cite{Peacock10,Hou12}, for example, have not been explicitly
presented. State of the art full-relativistic calculations of the
cubic $HfO_2$ carrier effective masses and frequency-dependent
dielectric function are presented in this work. The calculated
$\Gamma$-isotropic heavy and light electron effective masses are
several times heavier than the electron tunneling effective mass
measured recently \cite{Zhu13,Yeo14}. The calculated frequency
behavior of the imaginary part of $\varepsilon_{HfO_2}$ in the low
energy range is in good agreement with the recent measurements of
Lim {\it et al.} \cite{Lim13} by far ultraviolet (UV)
spectroscopic ellipsometry as well as with those of Edwards
\cite{Edwards} and of Schaeffer {\it et al} \cite{Edwards2}.

The {\it ab initio} all-electron self-consistent linear augmented
plane-wave (FLAPW) method, within the local-density functional
formalism (LDA) and the generalized gradient approximation (GGA),
is employed considering both non-- and full--relativistic
contributions to the band structure \cite{LAPW1}. The 5s--, 5p--,
4f--, 5d--, 6s--Hf; 2s-- and 2p--O electrons were treated as part
of the valence--band states. Cubic fluorite-type HfO$_2$  belongs
to the {\it Fm}3{\it m} ({\it Oh}) space (point) group, with Hf in
$(0,0,0)$, O in $({\pm}1/4,{\pm}1/4,{\pm}1/4)$, for which the
Brillouin zone (BZ) is a 14 face polyhedral \cite{Wyckoff,Wang18}.
Through a total energy minimization process within GGA, a lattice
constant $a=5.16$\thinspace\AA\ was obtained for cubic HfO$_2$.

Our calculated value $a_{min}=5.16$\thinspace\AA\ is close to
$a=5.06$\thinspace\AA\ and $a=5.04$\thinspace\AA\ obtained by
Fiorentini and Gulleri \cite{Fiorentini8} and Demkov
\cite{Demkov11}, respectively; $a=5.037$\thinspace\AA\ (LDA) and
$a=5.248$\thinspace\AA\ (GGA) calculated by Zhao and Vanderbilt
\cite{Vanderbilt7}. Note that the measured lattice constant
$a_{exp}$ is in the 5.08-5.30\thinspace\AA\ range according to the
data of Wang {\it et al.} \cite{Wang18}.

Figure 1 presents the band structure of cubic $HfO_2$ along
high-symmetry directions in the BZ and the total density of states
(TDOS), which were calculated both in the non-relativistic (dashed
lines) and relativistic (solid lines) approaches. The most
relevant contributions of the relativistic corrections are due to
the $Hf(p)$- and $(d)$-derived states, which contribute to the low
energy pattern (between -10 and -18 eV) of the band structure. A
remarkable shift, of $\approx 6$~eV upwards is seen for the
$Hf(p)$-derived states due to mass-velocity relativistic
contributions. The upward shift of the $Hf(p)$ states is followed
by a spin-orbit splitting of about $2$~eV which occurs between
$-10$ and $-12$~eV. The top of the valence band is mostly of
$O(p)$ character with some mixture with the $Hf(d)$- as well as
$Hf(f)$- derived states. $HfO_2$ exhibits a direct gap at the
$X$-point of $3.65\thinspace eV$. Demkov \cite{Demkov11} and
Peacock and Robertson \cite{Peacock10} obtained $3.4\thinspace eV$
using a plane wave basis set and nonlocal ultrasoft
pseudopotentials (CASTEP code) within the LDA. Our value for the
 band gap energy  is  in good agreement with the value of $3.6\thinspace eV$
 obtained  by Boer and Groot \cite{Boer9}
 via LAPW-LDA calculations within a relativistic approach but
 not including spin-orbit interaction effects. All theoretical values
  for the band gap energy are smaller
 than experimental, which is due to the well-known
 underestimation of  the energy values of conduction band
states in  {\it ab initio}  calculations within the local-density
functional theory. Through UV ellipsometry spectroscopy
measurements in bulk yttria--stabilized $HfO_2$ crystals  Lim {\it
et al.} \cite{Lim13} have assigned a value of $\sim 5.8\thinspace
eV$ to the $O(p)-Hf(d)$ band gap. We  note that a value of ~3.29
~eV is obtained from the non-relativistic calculation, and the
band gap is of an indirect nature in this case.

Table I presents the carrier effective masses in the $[100]$ and
$[111]$  ($\Gamma \rightarrow L$, $\Gamma \rightarrow X$, $X
\rightarrow \Gamma$) directions, calculated within and without
(see the values in parenthesis) the full-relativistic approach. In
the $X \rightarrow  W$ direction, the effective masses values are
too high to be evaluated within a parabolic fit. In both the
valence and conduction bands, the carrier effective masses are
demonstrated to be highly anisotropic. Relativistic effects are
seen to be mostly important in the $\Gamma-L$ direction for
electrons and holes, and in the $\Gamma-X$ direction in the case
of holes. The $\Gamma$-isotropic heavy and light electron
effective masses, defined as $m^{*}_{iso}=(8m^{*}_{\Gamma-L} +
6m^{*}_{\Gamma-X}){/ 14}$, have values $m^*_{he}=1.350$\thinspace
$m_o$ and $m^*_{le}=0.714$\thinspace $m_o$, respectively, where
 $m_0$ is the free electron mass. Both $m^*_{he}$ and $m^*_{le}$
are several times heavier than the appropriately named tunnelling
effective mass, which was recently estimated as:
$m^*_{e}=0.1$\thinspace $m_0$, through measurements of the
temperature dependence of gate leakage current and Fowler-Nordheim
tunnelling characteristics in metal/hafnium oxide/silicon
structures \cite{Zhu13}; and $m^*_{e}=0.17$\thinspace $m_0$,
through measurements of the direct tunnelling leakage current in
$n^+$ poly-Si gate NMOSFET with $HfO_2$ as gate dielectric
\cite{Yeo14}. The severe discrepancy between the calculated
$\Gamma$-isotropic electron effective masses and the experimental
tunnelling effective mass in $HfO_2$ highlights the limitations in
taking the latter as a measure of the bulk $HfO_2$
$\Gamma$-isotropic electron effective mass. The large discrepancy
may be related to the thin film characteristics and imperfections
of the $HfO_2$ gate dielectrics, as well as the influence of
$HfO_2/Si$ and $HfO_2/$metal interfaces on the gate current
density in the devices. Measurements of the carrier effective
masses in cubic $HfO_2$ samples have not been performed yet, which
precludes direct comparison of our theoretical estimates of the
carrier effective masses with experiment.

The imaginary part, $\epsilon_2(\omega)$, of the complex
dielectric function was obtained directly from full-- and
non--relativistic FLAPW electronic structure calculations, while
the Kramers-Kronig relation was used to obtain the real part,
$\epsilon_1(\omega)$. They are depicted in Fig. 2 over the energy
range 0 -- 16\thinspace eV. Relativistic contributions to the real
and imaginary parts of the dielectric constant are found to be
important, being responsible for a considerable shift of the main
peaks towards higher energies. Moreover, due to the relativistic
effects  a more detailed structure is seen in particular for
$\epsilon_2(\omega)$.  Figure 3 shows good agreement of the
calculated $\epsilon_2(\omega)$ with the experimental data
reported  by Lim {\it et al.} \cite{Lim13} for energies smaller
than $9.5\thinspace$eV,  and also with those performed by Edwards
\cite{Edwards} and by Schaeffer {\it et al} \cite{Edwards2}, all
using UV ellipsometry spectroscopy. For this comparison the whole
calculated spectrum for $\epsilon_2$ was shifted to higher
energies, by matching its energy threshold to the experimental
value of the $HfO_2$ gap energy. \cite{Lim13}

In conclusion, we have studied the band structure  of cubic bulk
$HfO_2$ using first-principles calculations. Conduction- and
valence-band effective masses were obtained and shown to be highly
anisotropic. Relativistic effects are shown to play an important
role, reflected  in the effective mass values and in the detailed
structure of the dielectric function. The $\Gamma$-isotropic heavy
and light electron effective masses were determined to be several
times heavier than the electron tunneling effective mass measured
recently \cite{Zhu13,Yeo14}. The calculated imaginary part of the
dielectric function was shown to agree well with experimental
measurements for energies smaller than $9.5\thinspace$eV
\cite{Lim13,Edwards,Edwards2}.







\vspace{0.8cm} \noindent The authors acknowledge the financial
support received from FAPESP and the Brazilian National Research
Council (CNPq) under contract NanoSemiMat/CNPq \# 550.015/01-9. We
thank Dr. A. Donegan for a critical reading of the manuscript.


\newpage

\begin{small}
\begin{figure}
\includegraphics[scale=0.6,clip=true]{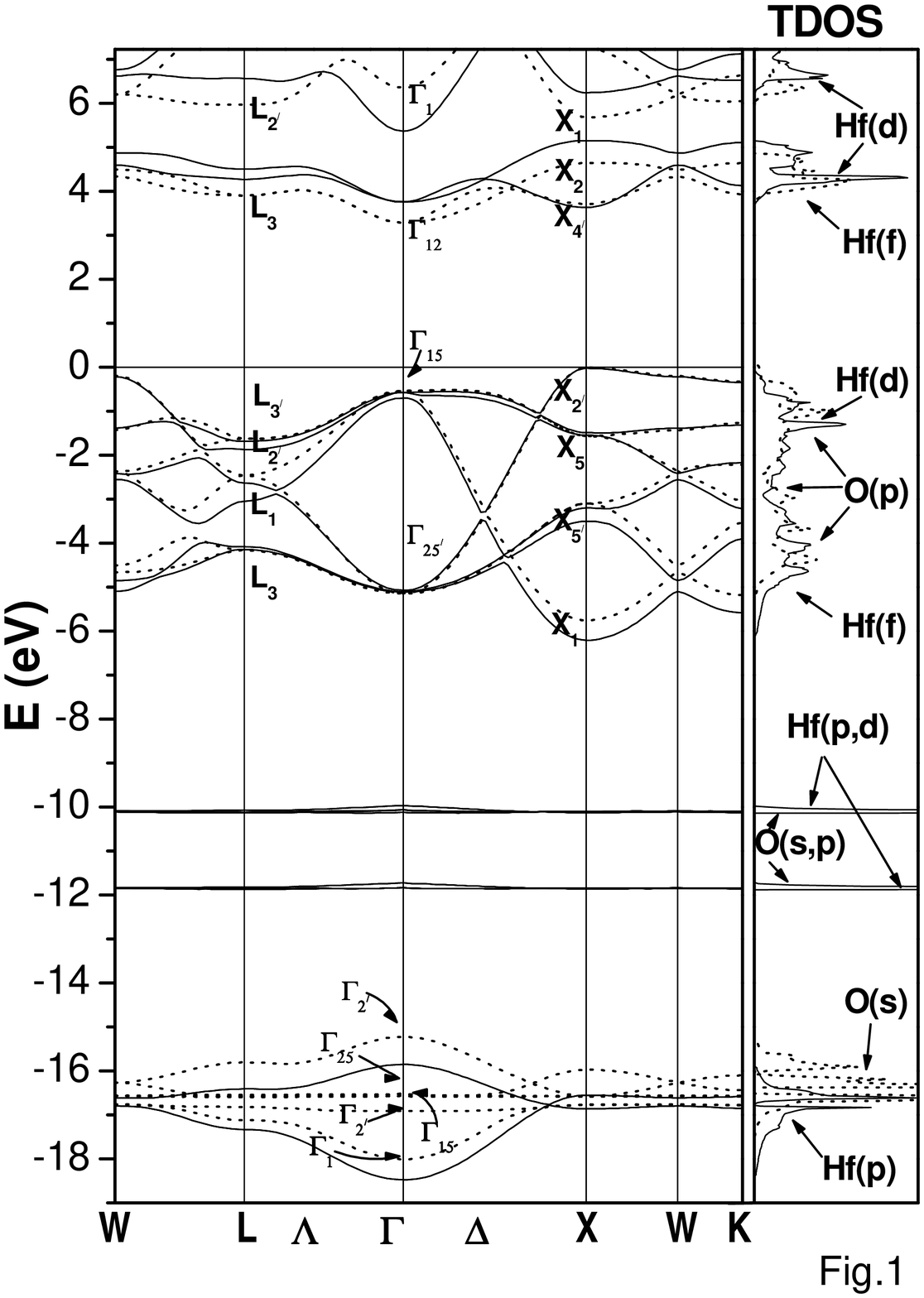}
\caption{Band structure of $HfO_2$ along high-symmetry axis of the
cubic BZ. The zero of energy (solid horizontal line) was set at
 the valence band maximum ($X$-point) for both, the
 full-relativistic and non-relativistic
 calculations.  The total density of states
(TDOS) is presented in the righthand column of the figure; the
main contributions for the DOS peaks are depicted. Full-(non-)
relativistic calculations are shown by solid (dashed) lines.
High-symmetry points of some bands were labeled according to their
irreducible representations given by group theory.} \label{Fig1}
\end{figure}
\end{small}

\begin{small}
\begin{figure}
\includegraphics[scale=0.6,clip=true]{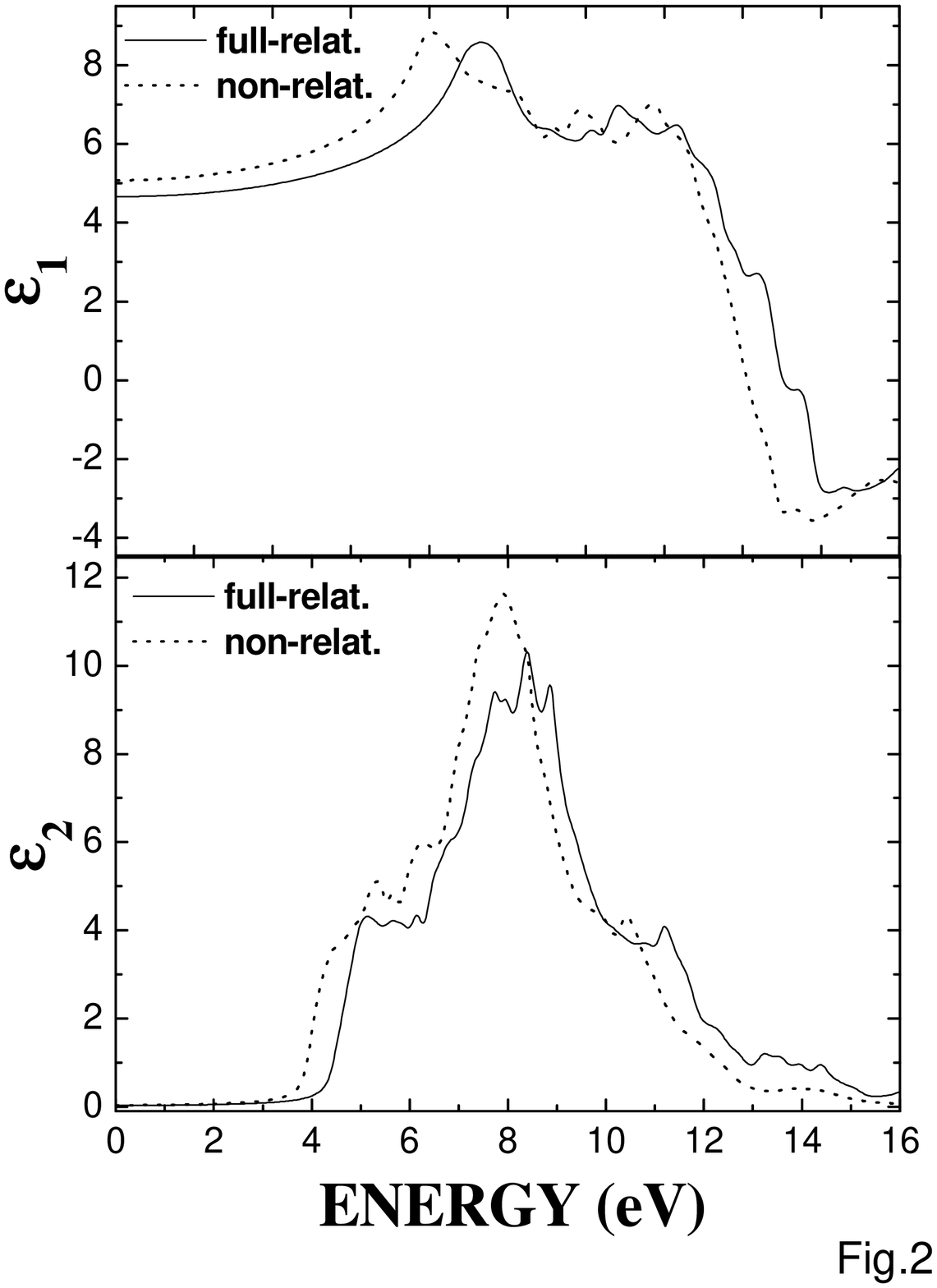}
\caption{Real ($\epsilon_1$) and imaginary ($\epsilon_2$) parts of
the $HfO_2$ complex dielectric function calculated within the
full--relativistic (solid line) and non--relativistic (dashed
line) approaches. The curves were plotted using a Lorentzian
broadening of 0.1 eV.} \label{Fig2}
\end{figure}
\end{small}


\begin{small}
\begin{figure}
\includegraphics[scale=0.6,clip=true]{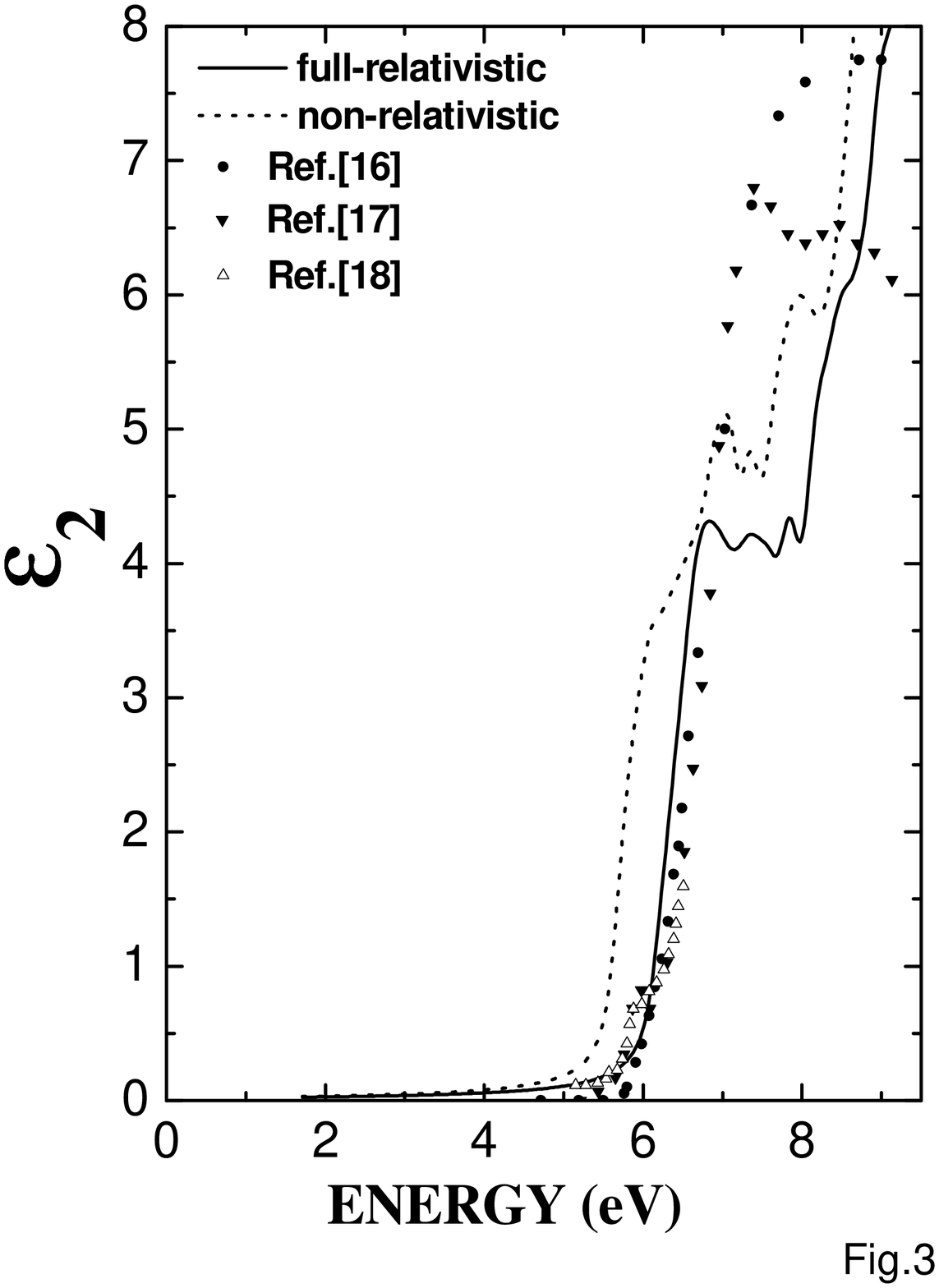}
\caption{Imaginary ($\epsilon_2$) part of the $HfO_2$ complex
dielectric function calculated within the full--relativistic
(solid line) and non--relativistic (dashed line) approaches. The
 calculated spectrum for $\epsilon_2$ was shifted to higher
energies, by matching its energy threshold  to the experimental
value of the $HfO_2$ gap energy as obtained in Ref.
[\onlinecite{Lim13}]. The experimental data of Lim {\it et al.}
\cite{Lim13} are depicted by full-dots, and those due to Edwards
\cite{Edwards} and Schaeffer {\it et al} \cite{Edwards2} are
depicted by upsidedown full-triangles and  open-triangles,
respectively.} \label{Fig3}
\end{figure}
\end{small}

\newpage

\begin{table}
\caption{Valence- and conduction-band effective masses at relevant
symmetry points in the BZ of $HfO_2$. Values are in units of the
free electron mass, $m_{0}$. The numbers in parentheses correspond
to effective-mass values obtained from a non-relativistic
calculation. }
\begin{center}
\begin{tabular}{c|ccc|ccc} \hline \hline
   &   &  valence   &  &  &conduction\\
 k-direction  & $m^{*}_{hh}$ & $m^{*}_{lh}$  & $m^{*v}_{so}$ &
 $m^{*}_{he}$  & $m^{*}_{le}$ & \\ \hline
 $\Gamma\rightarrow L$  & 0.871  & 0.580 & 0.720 & 0.930 & 0.823 \\
    &   (0.815)  &  (0.657)  &  (...) &  (0.781) &  (0.781) \\
$\Gamma\rightarrow X$    &  9.9    & 0.338   & 0.710  &1.91 & 0.570 \\
   &   (8.327)  &  (0.240)  &  (...) &  (2.017) &   (0.493) \\
\hline
 $X\rightarrow \Gamma$  & 0.286  & ... & ... & 1.268 & ... \\
   &   (0.268)  &  (...)  &  (...) &  (1.832) &  (...)  \\
\hline \hline
\end{tabular}
\end{center}
\label{tabI}
\end{table}

\end{document}